\documentclass[12pt]{amsart}

\usepackage[all]{xy}
\usepackage{amsmath}
\usepackage{amsfonts}
\usepackage{amssymb}
\usepackage{amscd}
\usepackage{amsthm}
\usepackage{latexsym}
\usepackage{amsbsy}
\usepackage{graphicx}
\usepackage{epstopdf}

\input xypic

\usepackage{color}

\newtheorem{thm}{Theorem}[section]
\newtheorem{prop}[thm]{Proposition}
\newtheorem{cor}[thm]{Corollary}
\newtheorem{lem}[thm]{Lemma}

\newtheorem{rem}[thm]{Remark}

\numberwithin{equation}{section}

\def\bG{{\mathbb G}}

\def\bK{{\mathbb K}}
\def\bL{{\mathbb L}}

\def\A{{\mathbb A}}

\renewcommand{\H}{{\mathbb H}}

\renewcommand{\P}{{\mathbb P}}
\def\Q{{\mathbb Q}}
\def\R{{\mathbb R}}
\def\Z{{\mathbb Z}}
\def\K{{\mathbb K}}

\def\m{{\mathfrak m}}

\def\cP{{\mathcal P}}

\def\cT{{\mathcal T}}

\def\cV{{\mathcal V}}

\def\Tr{{\rm Tr}}

\title[Motives and Periods in Bianchi IX gravity models]
{Motives and Periods in Bianchi IX gravity models}
\author{Wentao Fan, Farzad Fathizadeh and Matilde Marcolli}
\address{Princeton University \\ USA}
\email{wentaof@princeton.edu}

\address{California Institute of Technology \\ USA \newline \indent
Swansea University \\ UK}
\email{farzadf@caltech.edu}

\address{California Institute of Technology \\ USA \newline \indent
Perimeter Institute for Theoretical Physics \\ Canada \newline \indent
University of Toronto \\ Canada}
\email{matilde@caltech.edu}

\begin{document}
\maketitle

\begin{abstract}
We show that, when considering the anisotropic scaling factors and their derivatives as affine variables, 
the coefficients of the heat kernel expansion of the Dirac--Laplacian on $SU(2)$ Bianchi IX metrics 
are algebro-geometric periods of motives of complements in affine spaces of unions of quadrics and hyperplanes.
We show that the motives are mixed Tate and we provide an explicit computation of their Grothendieck
classes. 
\end{abstract}


\section{Introduction}

In this paper we continue our investigation of arithmetic structures arising in models of
Euclidean gravity based on the spectral action functional of \cite{CCact}. In \cite{FMMotives}
we showed that the heat-kernel Seeley-deWitt coefficients for the Dirac-Laplacian of the
Robertson--Walker metrics can be expressed as periods of mixed Tate motives given by
affine complements of unions of quadrics and hyperplanes. In \cite{FFMRationality} we
proved a rationality result for the Seeley-deWitt coefficients of the Bianchi IX metrics,
which generalizes an analogous rationality result for the Robertson--Walker case
conjectured in \cite{CC} and proved in \cite{FGK}. In \cite{FFM2} we proved that, in the
case of the two-parameter family of \cite{BaKo} of Bianchi IX gravitational instantons, the
Seeley-deWitt coefficients for the Dirac-Laplacian are vector valued modular forms. In
the present paper, we extend the result of \cite{FMMotives} on mixed Tate motives and
periods in the heat-kernel expansion of the Robertson--Walker metrics to the case of
the Bianchi IX metrics. Although the argument used in \cite{FMMotives} for the 
Robertson--Walker case does not immediately apply to the anisotropic Bianchi IX
metrics, we provide a different parameterization of the integrals computing the 
Seeley-deWitt coefficients, for which we can derive a very similar statement about
expressing these integrals as periods of certain mixed Tate motives given by 
complements of unions of quadrics and hyperplanes. 

\smallskip

The Bianchi IX metrics play an important role in Euclidean quantum gravity
and quantum cosmology in the form of minisuperspace models in Hartle--Hawking 
gravity, see \cite{Fang}. In view of a similar approach to quantum cosmology
based on the spectral action, currently being developed (see \cite{Mar2}), it
is interesting to investigate what role of arithmetic structures will 
play in such gravity models.

\section{Bianchi IX metrics and Dirac operators}

We consider here $SU(2)$-Bianchi IX metrics of the form
\begin{equation} \label{Bianchimetric}
ds^2 = w_1(t) \,w_2(t)\, w_3(t) \, dt^2 + 
\frac{w_2(t) \,w_3(t)}{w_1(t)} \, \sigma_1^2 + 
\frac{w_3(t) \,w_1(t)}{w_2(t)} \, \sigma_2^2+ 
\frac{w_1(t) \,w_2(t)}{w_3(t)} \, \sigma_3^2,    
\end{equation}
where the $\sigma_i$ are left-invariant $1$-forms on $SU(2)$-orbits 
satisfying the relations 
\[
d\sigma_1 = \sigma_2 \wedge \sigma_3, 
\qquad d\sigma_2 = \sigma_3 \wedge \sigma_1, 
\qquad d\sigma_3 = \sigma_1 \wedge \sigma_2. 
\]
This metric can be written locally as 
$ds^2 = \sum g_{\mu \nu} \,dx^\mu dx^\nu$,  in the set of local coordinates 
$x=(x^{\mu})_{\mu=1, \dots, 4} = (t, \eta, \phi, \psi)$, where the 
3-dimensional sphere $\mathbb{S}^3 \simeq SU(2)$ is parametrized 
by the map 
\[
( \eta, \phi, \psi) 
\mapsto 
\left ( \cos(\eta/2) \,e^{i (\phi+\psi)/2},   
\sin(\eta/2) \, e^{i (\phi-\psi)/2}  \right ). 
\] 
Here the parameters have the ranges $0 \leq \eta < \pi, 
0 \leq \phi < 2 \pi, 0 \leq \psi < 4 \pi$. 
The local formula of the Dirac operator $D$ 
of the metric in this coordinate system 
and its pseudodifferential symbol $\sigma_D$ can be 
computed as in \cite{FFMRationality}.
Using the symbol one can locally write the action of 
$D$ on a spinor $s$ as 
\begin{eqnarray} \label{pseudodifferentialOp}
D s (x) 
&=& 
(2 \pi)^{-2} \int e^{i \,x \cdot \xi} \, \sigma(D)(x, \xi) \, \hat s (\xi) \, d\xi  
\nonumber \\
&=& 
(2 \pi)^{-4} \int \int e^{i \,(x-y) \cdot \xi} \, \sigma(D)(x, \xi) \, s (y) \, dy\, d\xi,  
\end{eqnarray}
where $\hat s$ denotes the component-wise Fourier transform of $s$. 
In this formula $\xi$ is in fact an element of the cotangent fibre at the point $x$, 
which is identified with  $\R^4$. We have
\[
\sigma(D)(x, \xi) = q_1(x, \xi) + q_0 (x, \xi), 
\]
\begin{eqnarray}  \label{symbolofDiracq1q0}
q_1(x, \xi) &=&-\frac{i  \gamma^2
   \sqrt{w_1} \left(\csc (\eta )
   \cos (\psi ) \left(\xi _4 \cos (\eta
   )-\xi _3\right)+\xi _2 \sin (\psi
   )\right)}{\sqrt{w_2}
   \sqrt{w_3}} 
\end{eqnarray}
\begin{eqnarray*}   
&&+\frac{i 
   \gamma^3 \sqrt{w_2} \left(\sin
   (\psi ) \left(\xi _3 \csc (\eta )-\xi _4
   \cot (\eta )\right) +\xi _2 \cos (\psi
   )\right)}{\sqrt{w_1}
   \sqrt{w_3}} +\frac{i 
   \gamma^1 \xi _1}{\sqrt{w_1}
   \sqrt{w_2}
   \sqrt{w_3}}+\frac{i 
   \gamma^4 \xi _4
   \sqrt{w_3}}{\sqrt{w_1}
   \sqrt{w_2}},  
\end{eqnarray*}

\begin{eqnarray*}
q_0(x, \xi)&=&
\frac{1}{4\sqrt{w_{1}w_{2}w_{3}}}\left(\frac{w_{1}^{'}}{w_{1}}
+\frac{w_{2}^{'}}{w_{2}}+\frac{w_{3}^{'}}{w_{3}}\right)\gamma^{1} 
-\frac{\sqrt{w_{1}w_{2}w_{3}}}{4}\left(\frac{1}{w_{1}^{2}}
+\frac{1}{w_{2}^{2}}+\frac{1}{w_{3}^{2}}\right)
\gamma^{2}\gamma^{3} \gamma^4,
\end{eqnarray*}
where the $\gamma^i$ are $4 \times 4$ matrices such that 
$(\gamma^i)^2 = - I$ and $\gamma^i \gamma^j + \gamma^j \gamma^i = 0$ 
for $i \neq j$. 

\smallskip

Correspondingly, for the Dirac-Laplacian $D^2$ we have 
$$ \sigma(D^2)(x,\xi)=p_2(x,\xi)+p_1(x,\xi) +p_0(x,\xi), $$
where the homogeneous terms are given by
\begin{eqnarray}\label{p2p1p0eq}
p_2(x, \xi) 
&=&
q_1(x, \xi)^2, \nonumber \\
p_1(x, \xi)
&=& 
q_0(x, \xi) \, q_1(x, \xi)+ q_1(x, \xi) \, q_0(x, \xi)  \\ 
&& \qquad  \qquad + 
\sum _{j=1}^4 \left ( 
-i (\partial_{\xi_j} q_1) (x, \xi) \, (\partial_{x_j} q_1) (x, \xi) 
\right ) , \nonumber \\
p_0(x, \xi) &=& 
q_0(x, \xi)^2 +\sum _{j=1}^4 
\left (
-i (\partial_{\xi_j} q_1) (x, \xi) \, (\partial_{x_j} q_0) (x, \xi) 
\right ). \nonumber 
\end{eqnarray}
In particular, for later use, note that we can write the degree-two homogeneous term in the form
\[
p_2(x, \xi)=\left( \sum_{\mu, \nu=1}^4 g^{\mu \nu} \xi_\mu \xi_\nu \right) I .
\]

\section{Seeley-deWitt coefficients and periods}

The spectral action functional of Euclidean gravity, introduced in \cite{CCact}, is defined as a trace 
$\Tr(f(D/\Lambda))$ of the Dirac operator regularized by an even rapidly decaying function $f$
approximating a cutoff function on the Dirac spectrum, with $\Lambda$ an energy scale. It can be viewed
as a modified gravity model, since the leading terms in the large $\Lambda$ expansion include the 
Einstein-Hilbert action of gravity with cosmological term, as well as some higher derivative terms that 
include conformal gravity and Gauss--Bonnet gravity. Overviews of applications of the spectral
action functional to cosmology and particle physics can be found in \cite{Mar2} and \cite{WvS}.

\smallskip

The Seeley-deWitt coefficients of the heat-kernel expansion of the Dirac-Laplacian
$$ \Tr(e^{-s D^2}) \sim_{s\to 0^+} s^{-\dim(M)/2} \sum_{n=0}^\infty a_{2n}(D^2) s^n $$
determine the coefficients of the large energy asymptotic expansion of the spectral action functional, 
see \cite{CCact} and \S 1 of \cite{CoMa} for more details.
Thus, our approach to investigating the arithmetic properties of the spectral action models of gravity is based
on identifying arithmetic structures in the Seeley-deWitt coefficients of the heat-kernel
expansion of the Dirac-Laplacian. 

\smallskip
\subsection{The Seeley-deWitt coefficients as residues}

For any $n \in \Z_{\geq 1}$, the Seeley-deWitt coefficients $a_{2n}$ can be
computed as a noncommutative residue (see \cite{FFMRationality})
\begin{equation} \label{HeatCoefResidue}
a_{2n} = \frac{1}{32 \, \pi^{n+3}} \textnormal{Res}(\Delta_{2n}^{-1}),
\end{equation}
where 
\[
\Delta_{2n} = D^2 \otimes 1 + 1 \otimes \Delta_{\mathbb{T}^{2n-2}},
\]
with $\Delta_{\mathbb{T}^{2n-2}}$ the Laplacian of the 
flat metric on an auxiliary $(2n-2)$-dimensional torus 
$\mathbb{T}^{2n-2} = \left ( \mathbb{R}/\mathbb{Z} \right )^{2n-2}$. 
Since the operator $\Delta_{2n}$ is acting on the smooth sections of 
a vector bundle on a (2n+2)-dimensional manifold, in order to 
calculate $ \textnormal{Res}(\Delta_{2n}^{-1})$, we need the term 
that is positively homogeneous of order $-2n-2$ in the asymptotic expansion 
of the symbol of $\Delta_{2n}^{-1}$. We write 
\[
\sigma(\Delta_{2n}^{-1}) (x, \xi)
\sim_{\xi \to \infty}  
\sum_{m=-\infty}^{-2} \sigma_{m}(\Delta_{2n}^{-1})(x, \xi), 
\]
where each $\sigma_{m}(\Delta_{2n}^{-1})$ is (positively) homogeneous 
of order $m$ in $\xi$. 

\smallskip

By definition (see \cite{Wod1}, \cite{Wod2}) 
\begin{equation} \label{WodResDensity}
\text{Res} \left ( \Delta_{2n}^{-1} \right ) 
= 
\int_{M \times \mathbb{T}^{2n-2}}
\left ( \int_{|\xi |=1} 
\textrm{tr} \left (\sigma_{-2n-2}(x, \xi)  \right ) |\sigma_{\xi, \, 2n+1} | 
\right ) |dx^1 \wedge \cdots \wedge dx^{2n+2}|, 
\end{equation}
in which $\sigma_{\xi, \, 2n+1} $ is the volume form of the unit sphere 
$|\xi|=1$ in the cotangent fibre $\mathbb{R}^{2n+2} \simeq T_x^* (M \times \mathbb{T}^{2n-2})$, given by 
\begin{equation}\label{sigmaVol}
\sigma_{\xi, \, 2n+1} = \sum_{j=1}^{2n+2} (-1)^{j-1} \xi_j \, d\xi_1 
\wedge \cdots \wedge {\widehat d \xi_j} \wedge \cdots \wedge d \xi_{2n+2}. 
\end{equation}

\begin{rem} \label{homologouscyclesremark}{\rm 
Because of the homogeneity degree of $\sigma_{-2n-2}(x, \xi)$ in \eqref{WodResDensity} 
and the Stokes theorem, 
the integration over the sphere $|\xi|=1$ can be replaced with integration over the unit 
sphere of the metric or any other similar locus that is homologous to the sphere as a closed 
cycle, see Proposition~7.3 on page~265 of \cite{GVFbook}. }
\end{rem}

\smallskip 

The $\sigma_{m}(\Delta_{2n}^{-1})$ satisfy the recursive relations 
(see \cite{FFMRationality}) 
\begin{equation} \label{order-2}
\sigma_{-2}(\Delta_{2n}^{-1}) \left (x, \xi \right ) 
=  \left  ( p_2(x, \xi_1, \dots, \xi_4) +  \left (\xi_5^2 + \cdots + \xi_{2n+2}^2 \right ) I  \right )^{-1}, 
\end{equation}
and, for $m \leq -3$,
\begin{eqnarray} \label{-2-nterm} 
\sigma_{m}(\Delta_{2n}^{-1}) \left (x, \xi \right) =
\end{eqnarray}
 \[
- \left ( \sum_{\substack{ \alpha_1, \alpha_2, \alpha_4 \in \mathbb{Z}_{\geq 0}  \\ m < j \leq -2, \,\, \,\,\,\,0 \leq k \leq 2 
 \\j - \alpha_1 - \alpha_2 - \alpha_4 + k = m+2 }} 
 \frac{(-i)^{\alpha_1 + \alpha_2 + \alpha_4}}{\alpha_1! \,\alpha_2! \, \alpha_4! } 
\left (\partial_{\xi_1}^{\alpha_1 } \partial_{\xi_2}^{\alpha_2 }  \partial_{\xi_4}^{\alpha_4 }  \sigma_{j}(\Delta_{2n}^{-1}) \right ) 
\left ( \partial_t^{\alpha_1}  \partial_\eta^{\alpha_2}  \partial_\psi^{\alpha_4}  p_k \right ) \right ) 
\sigma_{-2}(\Delta_{2n}^{-1}). 
\]
Note that in this expression we have considered the fact that the symbol of the Dirac operator 
$D$ given by \eqref{symbolofDiracq1q0} is independent of the coordinate $\phi$.

\smallskip
\subsection{Seeley-deWitt coefficients as period integrals}

We focus here on the Seeley-deWitt coefficient before time integration, treating the
anisotropy coefficients $w_i$ and their derivatives as affine variables. We show that
for algebraic values of these variables the resulting coefficient is a period integral in
the algebro-geometric sense (see \cite{KoZa}), that is, an integral of an algebraic
differential form on a semi-algebraic set in an algebraic variety. 

\begin{rem}\label{alpha2n}{\rm 
In the following we use the notation $\alpha_{2n}$ for the Seeley-deWitt coefficient 
prior to integration in the time variable, namely
\begin{equation}\label{SdWt}
 a_{2n} = \int \alpha_{2n}(t) \, dt, 
\end{equation} 
where the $t$-dependence of $\alpha_{2n}$ is through the cosmic expansion factors (anisotropy
coefficients) $w_i(t)$ of the Bianchi IX metric, for $i=1,2,3$, and their derivatives, 
\begin{equation}\label{alpha2nwi}
 \alpha_{2n}(t)=\alpha_{2n}(w_i(t),w_i'(t),w_i''(t),\ldots, w_i^{(2n)}(t)). 
\end{equation}  }
\end{rem}

\smallskip

\begin{prop}\label{newcoords}
Introducing new variables 
\begin{equation}\label{Wicoeffs}\begin{array}{cc}
W_1=\frac{1}{\sqrt{w_1(t)} \sqrt{w_2(t)} \sqrt{w_3(t)}}, &
W_2=-\frac{\sqrt{w_1(t)}}{\sqrt{w_2(t)} \sqrt{w_3(t)}},  \\[3mm]
W_3=\frac{\sqrt{w_2(t)}}{\sqrt{w_1(t)} \sqrt{w_3(t)}},  &
W_4=\frac{\sqrt{w_3(t)}}{\sqrt{w_1(t)} \sqrt{w_2(t)}}
\end{array}\end{equation}
and the change of coordinates 
\[
\zeta_1 = \xi_1, 
\]
\[
\zeta_2= \xi _4 \cot (\eta ) \cos (\psi )-\xi _3 \csc (\eta ) \cos (\psi )+\xi _2 \sin (\psi ), 
\]
\[
\zeta_3 = -\xi _4 \cot (\eta ) \sin (\psi )+\xi _3 \csc (\eta ) \sin (\psi )+\xi _2 \cos (\psi )
\]
\[
\zeta_4 = \xi_4, \qquad \zeta_5 = \xi_5, \qquad \dots \qquad \zeta_{2n+2} = \xi_{2n+2}, 
\]
the expression  
$\textrm{tr} \left (\sigma_{-2n-2} \right )$  is given by 
\begin{eqnarray} \label{density-2n-2all}
\textrm{tr} \left (\sigma_{-2n-2}  \right )&=&\sum_{j=1}^{M_n} \Big \{  c_{j, 2n} \,
(\sin \eta)^{\beta_{0,1, j}} (\cos \eta )^{\beta_{0, 2, j}} \,
(\sin \psi )^{\beta_{1,1, j}} (\cos \psi)^{\beta_{1, 2, j}} \, \nonumber \\
&&
\qquad \qquad \qquad 
\frac{\zeta_1^{\beta_{1,j}} \zeta_2^{\beta_{2,j}}  \cdots \zeta_{2n+2}^{\beta_{2n+2,j}}  }
{Q_{W,2n}^{\rho_{j, 2n}}} 
\prod_{i=1}^{3} \omega_{i,0}^{k_{i,0,j}} \omega_{i, 1}^{k_{i,1,j}} \cdots \omega_{i, 2n}^{k_{i,2n,j}} \Big \},
\nonumber \\
\end{eqnarray}
where
\begin{eqnarray*}
&& c_{j, 2n} \in \Q, \\ 
&&\beta_{0, 1, j}, \beta_{0, 2, j}, \beta_{1, 1, j}, \beta_{1, 2, j}, k_{i, 0, j} \in \Z, \\
&&\beta_{1, j}, \dots, \beta_{2n+2, j}, \rho_{j, 2n}, k_{i, 1, j}, \dots, k_{i, 2n, j}  \in \Z_{\geq 0}, 
\end{eqnarray*}
where
\begin{equation}\label{qformQW2n}
Q_{W,2n}(\zeta_1,\ldots,\zeta_{2n+2}) = 
W_1^2 \zeta _1^2 + W_2^2 \zeta _2^2 + W_3^2 \zeta _3^2 + W_4^2 \zeta _4^2  + 
\zeta_5^2 + \cdots + \zeta_{2n+2}^2,
\end{equation}
with the variables $\omega_{i,j}$ associated with the cosmic 
expansion factors $w_1(t), w_2(t), w_3(t)$ given by 
\begin{equation}\label{omegaij}
\omega_{i,0} = w_i(t), \qquad \omega_{i, 1} = w'_i(t), \qquad \ldots \qquad \omega_{i, 2n} = w^{(2n)}_i(t) .
\end{equation}
\end{prop}

\proof   
This is a direct consequence of \eqref{order-2}, \eqref{-2-nterm}, the 
explicit formulas provided in \cite{FFMRationality} for the homogeneous 
symbols $p_2, p_1, p_0$ (which were calculated using \eqref{p2p1p0eq}), 
and the fact that 
\[
p_2(x, \xi_1, \dots, \xi_4) +  \xi_5^2 + \cdots + \xi_{2n+2}^2 
= 
Q_{W,2n}(\zeta_1,\ldots,\zeta_{2n+2}).
\] 

\endproof

We can then compute the Seeley-deWitt coefficient $\alpha_{2n}$ of \eqref{SdWt} as follows.

\begin{prop}\label{propSdW}
The Seeley-deWitt coefficient is given by the integral
\begin{equation}\label{SdWhom}
\alpha_{2n} =  \frac{1}{ \, \pi^{n+2}} \int_{0}^{\pi/2} \sin(\eta)\,d\eta 
 \int_{0}^{\pi/2} d\psi \int_{\sum_{i=1}^{2n+2} \zeta_i^2=1} 
\textrm{tr} \left (\sigma_{-2n-2}  \right ) \sigma_{\zeta, \, 2n+1} .
\end{equation}
\end{prop}

\proof By \eqref{WodResDensity}  and Remark \ref{homologouscyclesremark} we have
\begin{eqnarray} \label{a2nwithordinarysphere}
\alpha_{2n} 
&=&  \frac{1}{32 \, \pi^{n+3}} \int_{0}^{\pi} d\eta \int_{0}^{2 \pi} d\phi \int_{0}^{4 \pi} d\psi \int_{|\xi|_g=1} 
\textrm{tr} \left (\sigma_{-2n-2}  \right ) \sigma_{\xi, \, 2n+1} \nonumber \\
&=&  \frac{1}{ \, \pi^{n+2}} \int_{0}^{\pi/2} d\eta  \int_{0}^{\pi/2} d\psi \int_{|\xi|_g=1} 
\textrm{tr} \left (\sigma_{-2n-2}  \right ) \sigma_{\xi, \, 2n+1} ,
\end{eqnarray}
where $|\xi|_g =  \sum_{\mu, \nu =1}^4 g^{\mu \nu} \xi_\mu \xi_\nu + \xi_5^2 + \cdots + \xi_{2n+2}^2$.
Note that for the second identity in \eqref{a2nwithordinarysphere}, 
we used the fact that 
\[
\frac{1}{\sin (\eta )\, w_1(t) \, w_2(t) \,w_3(t)} \, 
\int_{|\xi|_g=1}\textrm{tr}  \left (\sigma_{-2n-2} \right ) \, 
\sigma_{\xi, \,2n+1} 
\]
is independent of the variables $\eta, \phi, \psi$. This fact is indeed associated with the 
symmetries of the Bianchi IX metric and was proved in \cite{FFMRationality}. 
Next observe that the
sphere $|\xi|_g=1$ determined by the metric $g$ is homologous to the 
sphere defined by 
\[
\sum_{i=1}^{2n+2} \zeta_i^2 = 
\xi _1^2+\xi _2^2+ \csc ^2(\eta ) \xi _3^2 +\csc ^2(\eta ) \xi _4^2 
-2 \cot (\eta ) \csc (\eta )  \xi _3 \xi _4 + \xi_5^2 + \cdots + \xi_{2n+2}^2=1 , 
\]
since the matrix 
\[
\left(
\begin{array}{cccccccc}
 1 & 0 & 0 & 0 & 0 & 0 & \cdots & 0 \\
 0 & 1 & 0 & 0  & 0 & 0&\cdots & 0 \\
 0 & 0 & \csc ^2(\eta ) & -\cot (\eta ) \csc (\eta ) &0 & 0&\cdots & 0 \\
 0 & 0 & -\cot (\eta ) \csc (\eta ) & \csc ^2(\eta ) &0 & 0&\cdots & 0 \\
 0 & 0 & 0 & 0 & 1 & 0& \cdots & 0 \\
 0 & 0 & 0 & 0 &0& 1 & \cdots & 0 \\
 \vdots & \vdots & \vdots & \vdots & \vdots &  & \ddots & \vdots \\
 0 & 0 & 0 & 0& 0 &  0& \cdots & 1
\end{array}
\right)
\]
is positive definite. By direct calculations one can also see that in the $\zeta$ 
coordinates one has
\[
\sigma_{\xi, \, 2n+1} = \sum_{j=1}^{2n+2} (-1)^{j-1} \xi_j \, d\xi_1 
\wedge \cdots \wedge {\widehat d \xi_j} \wedge \cdots \wedge d \xi_{2n+2} =
\]
\[
= \sin (\eta) \sum_{j=1}^{2n+2} (-1)^{j-1} \zeta_j \, 
d\zeta_1 
\wedge \cdots \wedge {\widehat d \zeta_j} \wedge \cdots \wedge d \zeta_{2n+2}
\]
\[
= \sin (\eta) \, \, \sigma_{\zeta, \, 2n+1}
\]
Therefore, considering Remark \ref{homologouscyclesremark}, we can write the Seeley-deWitt coefficient in the form \eqref{SdWhom}.
\endproof

Moreover, we need the following observation for the purpose of our description
of the Seeley-deWitt coefficients as periods. 

\begin{lem}\label{evenexp}
Only the terms with $\beta_{0,1,j},  \beta_{0,2,j}, \beta_{1,1,j},  
\beta_{1,2,j}  \in 2\Z$ 
in \eqref{density-2n-2all} contribute non-trivially to the calculation of 
$\alpha_{2n}$ in \eqref{SdWhom}. 
\end{lem}

\proof This follows from the fact that the integral 
\[
\frac{1}{\sin(\eta)} \int_{|\xi|_g=1} \text{tr}(\sigma_{-2n-2}) \, \sigma_{\xi, \,2n+1} 
= 
\int_{Q_{2n}=1} \text{tr}(\sigma_{-2n-2}) \, \sigma_{\zeta, \, 2n+1}
\]
is independent of the variables $\eta$ and $\psi$. Indeed, this implies that 
the terms in \eqref{density-2n-2all} where at least one of the integers $\beta_{0,1,j},  
\beta_{0,2,j}, \beta_{1,1,j},  \beta_{1,2,j}$ is odd cancel each other out 
after the integration over 
\[
Q_{2n} = \sum_{i=1}^4 W_i^2 \zeta_i^2 + 
\sum_{i=5}^{2n+2} \zeta_i^2 =1 .
\]
The terms where all the exponents are even, after the same integration, 
add up to an expression that is independent of the variables $\eta$ and 
$\psi$.  
\endproof

\smallskip

We introduce new coordinates, $\mu_1$ and 
$\mu_2$, defined by
\[
\mu_1 = - \cos(\eta) \cos(\psi), \qquad \mu_2 = \sin (\psi), 
\]
and we denote by $b_{-2n-2}$ the expression obtained from 
$\text{tr}(\sigma_{-2n-2})$ by removing all the terms for which at least one 
of the $\beta_{0,1,j},  \beta_{0,2,j}, \beta_{1,1,j},  \beta_{1,2,j}$  is 
an odd integer.  Our argument above shows that the following holds.

\smallskip

\begin{cor}\label{ratmuzeta}
The density $b_{-2n-2}$ is a rational 
expression in the variables $\mu_1$, $\mu_2$, $\zeta_1$, $\zeta_2$, $\ldots$, $\zeta_{2n+2}$ 
and in the affine variables $\omega_{i, j}$, $i \in \{ 1,2,3 \}$, $j \in \{ 1,2, \dots, 2n \}$ determined
by \eqref{omegaij}. 
\end{cor}

\proof This follows directly from the previous arguments and the identities
\[
\sin^2 (\psi) = \mu_2^2, \qquad \cos^2 (\psi) = 1- \mu_2^2, 
\]
\[
\sin^2(\eta) = \frac{1-\mu_1^2 - \mu_2^2}{1-\mu_2^2}, \qquad 
\cos^2(\eta) = \frac{\mu_1^2 }{1-\mu_2^2}. 
\]
\endproof

\smallskip

For the Seeley-deWitt coefficients this then gives the following expression as a
period in the algebro-geometric sense.

\begin{thm}\label{SdWmu}
For $\omega_{i,j}\in \bar\Q$, the Seeley-deWitt coefficient $\alpha_{2n}(\omega_{i,j})$ 
is a period in the algebro-geometric sense, given by the integral 
\begin{equation}\label{period}
\alpha_{2n} =  \frac{1}{ \, \pi^{n+2}} \int_{A_{2n}} 
\frac{ b_{-2n-2} }{1-\mu_2^2} \, d\mu_1 \wedge d\mu_2 \wedge \sigma_{\zeta, \, 2n+1}
\end{equation}
of an algebraic differential form 
$$ \frac{ b_{-2n-2} }{1-\mu_2^2} \, d\mu_1 \wedge d\mu_2 \wedge \sigma_{\zeta, \, 2n+1}, $$
defined on the complement in $\A^{2n+4}$ of the union of two hyperplanes
$$ H_\pm = \{ (\mu_1,\mu_2,\zeta_1,\ldots,\zeta_{2n})\in \A^{2n+4} \,:\, 
\mu_2 =\pm 1 \} $$
and the quadric defined by the vanishing of the quadratic form 
$Q_{W,2n}(\zeta_1,\ldots,\zeta_{2n})$, integrated over the semi-algebraic set
\begin{equation}\label{setA2n}
A_{2n} = \left \{  
(\mu_1, \mu_2, \zeta_1, \zeta_2, \dots, \zeta_{2n+2}) \in \A^{2n+4}(\R): 
0 < \mu_1, \mu_2 < 1 \text{ \,\, and \,\,} 
\sum_{i=1}^{2n+2} \zeta_i^2=1
\right \} .
\end{equation}
\end{thm}

\proof This follows from the previous results, using Corollary~\ref{ratmuzeta}
and the fact that 
\[
\sin(\eta) \, d\eta \wedge d\psi = \frac{1}{1-\mu_2^2} \, d\mu_1 \wedge d\mu_2 .
\]
By Proposition~\ref{newcoords}, the algebraic differential form $\frac{ b_{-2n-2} }{1-\mu_2^2}
\, d\mu_1 \wedge d\mu_2 \wedge \sigma_{\zeta, \, 2n+1}$ is defined on the complement in
$\A^{2n+4}$ of a hypersurface given by the union of two hyperplanes $H_\pm$ and
the quadric $\{ Q_{W,2n} =0 \}$.
\endproof

In the following section we describe the motives underlying these periods, and
we show that they are mixed Tate. 

\section{The motives}

The explicit computation of the motive can be obtained in a way that is similar to the
argument in the Robertson-Walker case of \cite{FMMotives}. Due to the different
choice of parameterization, the ambient space and the resulting motive is slightly
different, although the main result about the mixed Tate nature of the motive is unchanged.
The construction given here provides an alternative argument for the Robertson-Walker case
as a particular case. We treat the variables $W_i$ for $i=1,\ldots,4$ as parameters 
$W_i \in \bG_m(\K_i)$, where $\K_i$ are number fields. 

\smallskip

As in \cite{FMMotives} we adopt the following notation: we denote by $Z_{W,2n}\subset \P^{2n+1}$
the projective quadric determined by the quadratic form 
\begin{equation}\label{QW2n}
 Q_{W,2n} (\zeta_1, \ldots, \zeta_{2n+2}) =\sum_{i=1}^4 W_i^2 \zeta_i^2 +\sum_{i=5}^{2n+2} \zeta_i^2, 
\end{equation} 
for $W=(W_1,\ldots,W_4)\in \bG_m(\K)^4$,
$$ Z_{W,2n} =\{ (\zeta_1: \ldots: \zeta_{2n+2})\in \P^{2n+1}\,:\, Q_{W,2n}(\zeta_1, \ldots, \zeta_{2n+2}) =0 \}. $$
We also denote by $C^2 Z_{W,2n}$ the projective cone of $Z_{W,2n}$ in $\P^{2n+3}$ and we denote by
$\widehat Z_{W,2n}$ the affine cone in $\A^{2n+2}$ and by $\widehat{C^2Z}_{W,2n}$ the affine cone
of $C^2 Z_{W,2n}$ in $\A^{2n+4}$.

\smallskip

We are interested in the mixed motive 
\begin{equation}\label{mixmot}
 \m(\A^{2n+4}\smallsetminus (H_+ \cup H_- \cup \widehat{C^2Z}_{W,2n}), \Sigma), 
\end{equation}
where $H_\pm$ are the hyperplanes
\begin{equation}\label{Hpm}
H_\pm =\{ (\mu_1,\mu_2,\zeta_1, \ldots, \zeta_{2n+2})\in \A^{2n+4}\,:\, \mu_2=\pm 1 \}
\end{equation} 
and $\Sigma$ is the divisor in $\A^{2n+4}$ given by 
$$ \Sigma= \cup_{i=1}^2 \cup_{j=0}^1 H_{i,j}, $$
where $H_{i,j}$ are the hyperplanes
$$ H_{i,j}=\{ (\mu_1,\mu_2,\zeta_1, \ldots, \zeta_{2n+2})\in \A^{2n+4}\,:\, \mu_i =j \}. $$

\smallskip

We first give an explicit computation of the class 
\begin{equation}\label{GrClass}
 [\A^{2n+4}\smallsetminus (H_+ \cup H_- \cup \widehat{C^2Z}_{W,2n})] 
\end{equation} 
in the Grothendieck ring of varieties $K_0(\cV_\bL)$ with $\bL$ an extension
of $\K$, and then we prove that the motive \eqref{mixmot} is mixed Tate (as a
motive over $\bL$).

\smallskip
\subsection{The quadratic form and field extensions}\label{QchvarSec}
Let $\K$ be a number field that contains the fields $\K_i$, for $i=1,\ldots,4$ and $\Q(\sqrt{-1})$.
Over $\K$ consider the change of variables
\begin{equation}\label{varchangeQ}
\begin{array}{cc}
X_1 = W_1 \zeta_1 + i W_2 \zeta_2, & Y_1 = W_1 \zeta_1 - i W_2 \zeta_2 \\[2mm]
X_2 = i( W_3 \zeta_3 + i W_4 \zeta_4), & Y_2 = i(W_3 \zeta_3 - i W_4 \zeta_4) .
\end{array}
\end{equation}
In these variables the quadratic form $Q_{W,2}$ becomes the quadratic form
$$ X_1Y_1 - X_2Y_2 , $$
hence the projective quadric $Z_{W,2}\subset \P^3$ is the Segre quadric 
$$Z_{W,2}=\{ X_1Y_1 - X_2Y_2 =0 \} \simeq \P^1 \times \P^1. $$
Moreover, over the same field $\K$ the further changes of coordinates
\begin{equation}\label{varchangeQn}
X_n =\zeta_{2n-1} +i \zeta_{2n}, \ \ \ Y_n =\zeta_{2n-1}- i \zeta_{2n}
\end{equation}
transform the quadratic form $Q_{W,2n}$ into the form
\begin{equation}\label{indQ2n}
Q_{W,2n-2}(\zeta_1,\ldots,\zeta_{2n}) + X_n Y_n. 
\end{equation}

\smallskip
\subsection{The Grothendieck class}
The Grothendieck ring of varieties $K_0(\cV_\bL)$  is generated by isomorphism classes $[X]$
of varieties over $\bL$ with the inclusion-exclusion relation $[X]=[Y]+[X\smallsetminus Y]$
for closed subvarieties $Y\hookrightarrow X$, and the product relation $[X]\cdot [Y]=[X\times Y]$.
In order to compute the Grothedieck class \eqref{GrClass}, we use the following facts, which are
a variant of Lemma~4.1 of \cite{FMMotives}.

\begin{lem}\label{classcones}
Let $Z \subset \P^{2n+1}$ is a projective hypersurface and let $C^2 Z\subset \P^{2n+3}$,
$\hat Z\subset \A^{2n+2}$ and $\widehat{C^2 Z}\subset \A^{2n+4}$ be the projective and affine
cones as above. Also let $H_\pm$ be two hyperplanes in $\A^{2n+4}$ with $H_+\cap H_-=\emptyset$
and with intersections $H_\pm \cap \widehat{C^2 Z}$ given by sections of the cone. Then the
Grothendieck classes satisfy
\begin{itemize}
\item $[\A^{2n+4}\smallsetminus \widehat{C^2 Z}]= \bL^{2n+4} - \bL^3 [Z] + \bL^2 ( [Z] -1)$,
\item $[\A^{2n+4}\smallsetminus (\widehat{C^2 Z} \cup H_+ \cup H_-)]= \bL^{2n+4}- 2 \bL^{2n+3} -
\bL^3 [Z] +3 \bL^2 [Z]  - 2 \bL [Z]  - \bL^2  +2 \bL$.
\end{itemize}
\end{lem}

\proof Let $\bL=[\A^1]$ be the Lefschetz motive, the Grothendieck class of the affine line. 
We have $[\A^{2n+2}\smallsetminus \hat Z]=(\bL-1) [\P^{2n+1}\smallsetminus Z]$
since $[\hat Z]=(\bL-1)[Z]+1$. Moreover, we have $[CZ]=\bL [Z]+1$, since the projective
cone is the union of a copy of $Z$ and a copy of the affine cone $\hat Z$. Similary, we
have $[C^2 Z]=\bL [CZ]+1=\bL^2 [Z]+\bL +1$. We then have 
$$ [\A^{2n+4}\smallsetminus \widehat{C^2 Z}]=(\bL-1) [\P^{2n+3}\smallsetminus C^2 Z] =
\bL^{2n+4}-1 - (\bL-1) [C^2 Z] = $$ $$ \bL^{2n+4}-1 - (\bL-1) (\bL^2 [Z]+\bL +1) 
= \bL^{2n+4} - \bL^3 [Z] + \bL^2 ( [Z] -1). $$
By inclusion-exclusion we have 
$$ [\widehat{C^2 Z} \cup H_+ \cup H_-]=[\widehat{C^2 Z}]+[H_-\cup H_+]-[\widehat{C^2 Z} \cap (H_+ \cup H_-)]
= [\widehat{C^2 Z}]+ 2\bL^{2n+3} - 2[\widehat{CZ}], $$
where $[\widehat{C^2 Z}]=(\bL-1)[C^2Z]+1 =
\bL^3 [Z] +\bL^2 -\bL^2 [Z]  = \bL^3 [Z] -\bL^2 ([Z]-1)$ and 
$[\widehat{CZ}]=(\bL-1)[CZ]+1=\bL^2 [Z] -\bL( [Z] -1)$, so that we obtain
$$ [\widehat{C^2 Z} \cup H_+ \cup H_-]= \bL^3 [Z] -\bL^2 ([Z]-1) + 2 \bL^{2n+3} - 2 (\bL^2 [Z] -\bL( [Z] -1))  $$
$$ = 2 \bL^{2n+3} + \bL^3 [Z] -3 \bL^2 [Z]  + 2 \bL [Z]  + \bL^2 -2 \bL .$$
\endproof

\smallskip

\begin{prop}\label{classQ2n}
Let $Q_{W,2n}$ be the quadratic form \eqref{QW2n} and $Z_{W,2n}\subset \P^{2n+1}$ the
projective quadric defined by the vanishing of $Q_{W,2n}$. Let $C_{2n}=[\A^{2n+2}\smallsetminus \hat Z_{W,2n}]$
be the Grothendieck class in $K_0(\K)$ of the affine hypersurface complement. This is given by
\begin{equation}\label{C2nClass}
C_{2n} = \bL^{2n+2}- \bL^{2n+1} - \bL^{n+1} + \bL^n .
\end{equation}
Similarly, we have $[Z_{W,2n}]=1+\bL+\cdots +\bL^{n-1} + 2 \bL^n +\bL^{n+1}+\cdots +\bL^{2n}$
and 
\begin{equation}\label{affC2Zclass}
[\A^{2n+4}\smallsetminus \widehat{C^2 Z}_{W,2n}]=\bL^{2n+4} - \bL^{2n+3} -\bL^{n+3} +\bL^{n+2}.
\end{equation}
\begin{equation}\label{affC2ZclassH}
[\A^{2n+4}\smallsetminus (\widehat{C^2 Z}_{W,2n} \cup H_+ \cup H_-)] =
\bL^{2n+4} -3 \bL^{2n+3} + 2 \bL^{2n+2} - \bL^{n+3} +3 \bL^{n+2} -2 \bL^{n+1}.
\end{equation}
for $H_\pm=\{ \mu_2=\pm 1 \}\subset \A^{2n+4}$.
\end{prop}

\proof We first show that the classes $C_{2n}$ satisfy the recursive formula 
\begin{equation}\label{recC2n}
C_{2n} = \bL^{2n+2} -2 \bL^{2n+1} +\bL^{2n} + \bL C_{2n-2} .
\end{equation}
To see this, consider the condition that $Q_{W,2n} \neq 0$. By \eqref{indQ2n}, using the
change of variables \eqref{varchangeQn} over $\bK$, this is equivalent to
$$ Q_{W,2n-2}(\zeta_1,\ldots \zeta_{2n}) + X_n Y_n \neq 0. $$
Suppose $X_n=0$. Then $Y_n\in \A^1$ and $Q_{W,2n-2}(\zeta_1,\ldots \zeta_{2n})\neq 0$.
Thus, this case contributes a term $\bL \cdot C_{2n-2}$ to the class $C_{2n}$. The case
$X_n\neq 0$ gives $$Y_n \neq \frac{Q_{W,2n-2}(\zeta_1,\ldots \zeta_{2n})}{X_n}, $$
which gives $(\zeta_1,\ldots \zeta_{2n})\in \A^{2n}$ and $Y_n\in \bG_m$ with $X_n\in \bG_m$.
Thus, this case contributes a term $[\bG_m]^2 \bL^{2n}=\bL^{2n} (\bL-1)^2$. This gives 
$C_{2n}=\bL^{2n+2}- 2 \bL^{2n+1} +\bL^{2n}+\bL \cdot C_{2n-2}$. We can then verify \eqref{C2nClass}
by induction. When $n=1$, we know from \S \ref{QchvarSec} that the change of variables \eqref{varchangeQ}
over $\bK$ transforms the quadric $Q_{W,2}$ into the quadric $X_1Y_1 - X_2Y_2$, hence $[Z_{W,2}]
=[\P^1 \times \P^1]=\bL^2 + 2\bL +1$ in $K_0(\K)$. Thus we have $[\hat Z_{W,2}]=(\bL-1)[Z_{W,2}]+1=
(\bL-1)(\bL^2 + 2\bL +1)+1=\bL^3 +2 \bL^2 +\bL -\bL^2 -2 \bL -1 +1=\bL^3 + \bL^2 -\bL$,
hence $C_2=\bL^4 -  \bL^3 - \bL^2 +\bL$. Then suppose that 
$C_{2n-2}=\bL^{2n}-\bL^{2n-1}-\bL^n +\bL^{n-1}$. We obtain
$C_{2n} = \bL^{2n+2} -2 \bL^{2n+1} +\bL^{2n} + \bL (\bL^{2n}-\bL^{2n-1}-\bL^n +\bL^{n-1})
= \bL^{2n+2}  - \bL^{2n+1} - \bL^{n+1} + \bL^n$. We also have 
$[\hat Z_{W,2n}]=\bL^{2n+1} +\bL^{n+1} - \bL^n$ and $[Z_{W,2n}]=([\hat Z_{W,2n}]-1) (\bL-1)^{-1}=
(\bL^{2n+1} +\bL^{n+1} - \bL^n-1)  (\bL-1)^{-1}= 1+\bL+\cdots +\bL^{2n} + \bL^n$, hence
$$ [\A^{2n+4}\smallsetminus \widehat{C^2 Z}_{W,2n}]= \bL^{2n+4} - \bL^3 [Z_{W,2n}] + \bL^2 ( [Z_{W,2n}] -1) $$
$$ = \bL^{2n+4} - \bL^3 (1+\bL+\cdots +\bL^{2n} + \bL^n) + \bL^2 (1+\bL+\cdots +\bL^{2n} + \bL^n) - \bL^2 $$
$$ = \bL^{2n+4} - \bL^{2n+3}  -\bL^{n+3} +\bL^{n+2},  $$
as the other terms cancel in a telescopic sum. Similarly, we have
$$ [\A^{2n+4}\smallsetminus (\widehat{C^2 Z}_{W,2n} \cup H_+ \cup H_-)] = 
\bL^{2n+4}- 2 \bL^{2n+3} -\bL^3 [Z_{W,2n}] +3 \bL^2 [Z_{W,2n}] $$ $$ - 2 \bL [Z_{W,2n}]  - \bL^2  +2 \bL 
 = \bL^{2n+4}- 2 \bL^{2n+3} -\bL^3  (1+\bL+\cdots +\bL^{2n} + \bL^n) $$ 
 $$ + 3 \bL^2 (1+\bL+\cdots +\bL^{2n} + \bL^n)  - 2 \bL (1+\bL+\cdots +\bL^{2n} + \bL^n) -\bL^2 + 2 \bL $$
 $$ =\bL^{2n+4} -3 \bL^{2n+3} + 2 \bL^{2n+2} - \bL^{n+3} +3 \bL^{n+2} -2 \bL^{n+1}. $$
\endproof

\smallskip
\subsection{The mixed motive}

The result of Proposition~\ref{classQ2n} shows that the Grothendieck classes in $K_0(\K)$
of the complements $\A^{2n+4}\smallsetminus \widehat{C^2 Z}_{W,2n}$ and
$\A^{2n+4}\smallsetminus (\widehat{C^2 Z}_{W,2n} \cup H_+ \cup H_-)$ are in the Tate
subring $\Z[\bL] \subset K_0(\K)$. We now consider the mixed motive \eqref{mixmot}, as
an element in the Voevodsky triangulated category of mixed motives, \cite{Voe}, and we show that
it is in the triangulated subcategory of mixed Tate motives. The argument is analogous to
Theorem~4.3 and Proposition~4.5 of \cite{FMMotives}. We present it here explicitly for completeness. 

\begin{thm}\label{mixmothm}
The mixed motive $\m(\A^{2n+4}\smallsetminus (H_+ \cup H_- \cup \widehat{C^2Z}_{W,2n}), \Sigma)$
over the field $\K$ is a mixed Tate motive.
\end{thm}

\proof Using the change of variables \eqref{varchangeQn} we see that, over the field extension $\K$ the
quadratic form becomes isotropic, namely $Q_{W,2n}|_\K= (n+1) \H$, where $\H=\langle 1, -1 \rangle$ is the hyperbolic
quadratic form. This implies that, over $\K$, the motive $\m(Z_{W,2n})$ is given by (see \cite{Vishik2})
$$  \m(Z_{W,2n}) = \Z(n)[2n]\oplus \Z(n)[2n] \oplus \bigoplus_{k=0,\ldots, n-1,n+1,\ldots, 2n} \Z(k)[2k]. $$
this motivic decomposition of the motive corresponds to the expression  
$[Z_{W,2n}]=1+\bL+\cdots +\bL^{n-1} + 2 \bL^n +\bL^{n+1}+\cdots +\bL^{2n}$ for the Grothendieck class. 
The Gysin distinghuished triangle in the Voevodsky category gives 
$$ \m(\P^{2n+1}\smallsetminus Z_{W,2n}) \to \m(\P^{2n+1}) \to \m(Z_{W,2n})(1)[2] \to \m(\P^{2n+1}\smallsetminus Z_{W,2n})[1]. $$
Since two of the three terms, $\m(\P^{2n+1})$ and $\m(Z_{W,2n})(1)[2]$ are mixed Tate, the third term $\m(\P^{2n+1}\smallsetminus Z_{W,2n})$
is also mixed Tate. Note then that, when taking a projective cone, the map $\P^{2n+2}\smallsetminus CZ_{W,2n} \to \P^{2n+1}\smallsetminus Z_{W,2n}$ is an $\A^1$-fibration, and so is the map $\P^{2n+3}\smallsetminus C^2 Z_{W,2n} \to \P^{2n+2}\smallsetminus CZ_{W,2n}$.
By homotopy invariance of the Voevodsky motive, we then have $\m(\P^{2n+3}\smallsetminus C^2 Z_{W,2n})\simeq 
\m(\P^{2n+1}\smallsetminus Z_{W,2n})$. Thus, the motive $\m(\P^{2n+3}\smallsetminus C^2 Z_{W,2n})$ is also mixed Tate. 
The relation between the motive $\m(\P^{2n+3}\smallsetminus C^2 Z_{W,2n})$ and the motive $\m(\A^{2n+4}\smallsetminus \widehat{C^2 Z}_{W,2n})$ is obtained by considering the $\bG_m$-bundle $\cT=\A^{2n+4}\smallsetminus \widehat{C^2 Z}_{W,2n} \to \P^{2n+3}\smallsetminus C^2 Z_{W,2n}$ and the associated $\P^1$-bundle $\cP$ and the Gysin distinguished triangle of \cite{Voe}, p.197, 
$$ \m(\cT) \to \m(\cP) \to \m(\cP\smallsetminus \cT)^*(1)[2] \to \m(\cT)[1]. $$
The motive of a $\P^1$-bundle over a base $X$ satisfies $\m(\cP)=\m(X)\oplus \m(X)(1)[2]$, hence it is mixed Tate 
if $\m(X)$ is mixed Tate. The motive $\m(\cP\smallsetminus\cT)$ is mixed Tate because it consists of two copies of $X$.
Thus, in the above triangle both $\m(\cP)$ and $\m(\cP\smallsetminus\cT)$ are mixed Tate, hence the third term $\m(\cT)=\m(\A^{2n+4}\smallsetminus \widehat{C^2 Z}_{W,2n})$ is also mixed Tate. Next we show that the motive $\m(\A^{2n+4}\smallsetminus (\widehat{C^2 Z}_{W,2n} \cup H_+ \cup H_-))$ is mixed Tate as well. We use the Mayer-Vietoris distinguished triangle in the Voevodsky category
$$ \m(U\cap V) \to \m(U)\oplus \m(V) \to \m(U\cup V) \to \m(U\cap V) [1], $$
applied to the open sets $U=\A^{2n+4}\smallsetminus \widehat{C^2 Z}_{W,2n}$ and $V=\A^{2n+4}\smallsetminus (H_+\cup H_-)$,
with $U\cup V=\A^{2n+4}\smallsetminus (\widehat{C^2 Z}_{W,2n} \cap (H_+\cup H_-))$ and $U\cap V=
\A^{2n+4}\smallsetminus (\widehat{C^2 Z}_{W,2n} \cup H_+\cup H_-)$. We want to show that $\m(U\cap V)$ is
mixed Tate. By the Mayer-Vietoris triangle it suffices to show that $\m(U)$, $\m(V)$, and $\m(U\cup V)$ are all mixed Tate. 
We know that $\m(U)$ is mixed Tate by the previous argument. To see that $\m(V)$ is mixed Tate observe that $\m(H_+\cup H_-)$
certainly is, hence the Gysin triangle ensures that $\m(V)$ is also mixed Tate. In the case of 
$\m(U\cup V)$, the intersection $\widehat{C^2 Z}_{W,2n} \cap (H_+\cup H_-)$ consists of two sections of the cone,
isomorphic to $\widehat{CZ}_{W,2n}$, hence $\m(\widehat{C^2 Z}_{W,2n} \cap (H_+\cup H_-))=\m(\widehat{CZ}_{W,2n})\oplus
\m(\widehat{CZ}_{W,2n})$. The motive $\m(\widehat{CZ}_{W,2n})$ is mixed Tate because the motive of the complement
is by the previous argument about homotopy invariants and the Gysin triangle. Thus, the motive of the complement
$\m(\A^{2n+4}\smallsetminus (\widehat{C^2 Z}_{W,2n} \cap (H_+\cup H_-))$ is also mixed Tate, again by an application
of the Gysin triangle. Thus, by Mayer-Vietoris we have obtained that $\m(\A^{2n+4}\smallsetminus (\widehat{C^2 Z}_{W,2n} 
\cup H_+\cup H_-))$ is mixed Tate. Finally, the motive \eqref{mixmot} also fits in a distinguished triangle where
two of the terms, $\m(A^{2n+4}\smallsetminus (\widehat{C^2 Z}_{W,2n} 
\cup H_+\cup H_-))$ and $\m(\Sigma)$ are mixed Tate, hence it is also mixed Tate.
\endproof

\medskip

\begin{rem}\label{remQRost}{\rm Assuming for simplicity that $W_i\in \bG_m(\Q)$, the motive $\m(Z_{W,2n})$ over $\Q$, 
where the quadratic form $Q_{W,2n}$ is not isotropic, can be expressed in terms of ``forms of Tate motives",
which become Tate motives after passing to a field extension. These are the Rost motives of quadrics,
see \cite{Rost}, \cite{Vishik1}, \cite{Vishik2}, and \S 4.6 of \cite{FMMotives}.}
\end{rem}

\medskip

\begin{rem}\label{remModForms} {\rm In \cite{FFM2} we proved that, for Bianchi IX metrics that are gravitational
instantons (Einstein and self-dual), the heat kernel coefficients are vector valued modular forms. Thus, we see two
different arithmetic structures associated to these  heat kernel coefficients: as we have shown here, if one fixes
an algebraic value of the anisotropy coefficients $w_i$ (hence of the coefficients $W_i$), then the corresponding
Seeley-deWitt coefficients are periods of mixed Tate motives; on the other hand, if one considers the $w_i$
and an overall conformal factor $F$ as functions of the cosmological time $\mu$ and the two parameters
$(p,q)$ determining the family of solutions of the gravitational instanton equations, then the Seeley-deWitt
coefficients are  vector valued (meromorphic) modular forms in the variable $i \mu$ in the upper half plane.}
\end{rem}


\subsection*{Acknowledgement} The third author acknowledges 
support from NSF grant DMS-1707882 and from the Perimeter
Institute for Theoretical Physics.

\end{document}